\documentstyle[pre,aps,twocolumn,amssymb]{revtex}

\begin{document}

\draft
\title{Semiclassical Quantization by Harmonic Inversion: Comparison of
       Algorithms}
\author{Thomas Bartsch, J\"org Main, and G\"unter Wunner}
\address{Institut f\"ur Theoretische Physik 1, Universit\"at Stuttgart,
	 D-70550 Stuttgart, Germany}
\date{\today}
\maketitle

\begin{abstract}
Harmonic inversion techniques have been shown to be a powerful tool for
the semiclassical quantization and analysis of quantum spectra of both
classically integrable and chaotic dynamical systems.
Various computational procedures have been proposed for this purpose.
Our aim is to find out which  method is numerically most efficient.
To this end, we summarize and discuss the different techniques and compare 
their accuracies by way of two example systems.
\end{abstract}


\section{Introduction}
Semiclassical trace formulae relate the spectra of quantum systems to the
periodic orbits of the pertinent classical systems \cite{Gut90}.
They yield expansions of the quantum response function of the form
\begin{equation}
\label{spur}
   g(E)
 = \sum_k \frac{d_k}{E-E_k+\text{i}\epsilon}
 = \sum_{\text{po}} {\cal A}_{\text{po}}
     \text{e}^{\text{i} S_{\text{po}}/\hbar} \; .
\end{equation}
Here, $E_k$ are the energy eigenvalues of the quantum system, $d_k$ their
multiplicities, $S_{\text{po}}$ is the action of a classical periodic orbit,
${\cal A}_{\text{po}}$ an amplitude that can be calculated from classical
mechanics (including phase information given by the Maslov index), and 
the sum on the right-hand side extends over all periodic orbits and usually
diverges for real energies $E$. 
Thus, the quantal information cannot be extracted directly from the 
semiclassical expansion.

One particular and widely applicable method to overcome the convergence 
problems of the periodic orbit sum is semiclassical quantization by 
harmonic inversion \cite{Mai98,Mai99}.
By a Fourier transform of Eq.\ (\ref{spur}) the problem of semiclassical 
quantization can be recast as a harmonic inversion problem, viz.\ the 
extraction of the frequencies $\omega_k=E_k/\hbar$ and amplitudes $d_k$ 
from a given time signal
\begin{equation}
\label{hi}
  C(t) = \sum_k d_k \text{e}^{-\text{i} \omega_k t} \; .
\end{equation}
Especially intriguing, and important, are systems possessing a classical
scaling property, i.e., the classical dynamics does not depend on an
external scaling parameter $w$ and the action 
$S_{\text{po}}=ws_{\text{po}}$ of periodic orbits varies linearly with $w$,
with $s_{\text{po}}$ the scaled action.
This is not a severe restriction since it covers a variety of systems, 
such as systems with homogeneous potentials, billiard systems, or the 
hydrogen atom in static external fields.
For scaling systems the semiclassical signal which has to be harmonically 
inverted has the special form of a sum of $\delta$ functions with peaks at 
positions given by the scaled actions $s_{\text{po}}$ of the periodic orbits,
\begin{equation}
\label{C_sc}
  C(s) = \sum_{\rm po} {\cal A}_{\rm po} \delta\left(s-s_{\rm po}\right) \; .
\end{equation}
The frequencies of the signal (\ref{C_sc}) are the semiclassical approximations
to the quantum eigenvalues $w_k$ of the scaling parameter.
By the same token, harmonic inversion of signals with the functional form 
(\ref{C_sc}) also plays a key r\^ole in the high resolution analysis of the 
density of states $\varrho(E)=\sum_n\delta(E-E_n)$ of quantum spectra, in 
an effort to extract information about the underlying classical dynamics 
\cite{Mai99,Mai97,Gre00}.

In practical applications, the signal (\ref{C_sc}) is always known in 
a finite range $0 \le s \le S_{\text{max}}$ only.
The signal length $S_{\text{max}}$ is often fixed or at least hard to
increase, e.g., for periodic orbit quantization of classically chaotic
systems where the number of periodic orbits proliferates exponentially
with the signal length.
To obtain the optimum results from the harmonic inversion procedure,
it is essential to choose an algorithm which allows one to extract the
most precise estimates for the spectral parameters $\{\omega_k,d_k\}$
from the signal of a given length $S_{\text{max}}$.

Various computational procedures have been proposed for the harmonic inversion
of signals of the type (\ref{C_sc}).
However, a systematic study of the relative merits and demerits of the methods
and a quantitative study of their efficiencies is still lacking.
To remedy this situation, we summarize and discuss different techniques of 
harmonic inversion and compare their accuracies in the application to two 
simple albeit typical example systems for which exact trace formulae are known.
The aim is to pin down the numerically most efficient method for harmonic 
inversion.

\section{Harmonic inversion of delta function signals}
\label{methods:sec}
Due to the finite signal length $S_{\text{max}}$, the analysis of 
the signal by conventional Fourier transform is limited by the 
``uncertainty principle'', which states that in a signal of finite 
length $S_{\text{max}}$ two frequencies can only be resolved if their 
difference is larger than $2\pi/S_{\text{max}}$.
The uncertainty principle can be circumvented by the application of
high-resolution methods \cite{Wal95,Man97} which extract a discrete 
set of frequencies and amplitudes without imposing any {\em a priori} 
restrictions on the frequencies $\omega_k$.
However, an uncertainty remains in the weaker form of the ``informational 
uncertainty principle'' \cite{Man97}, which states that the signal length 
$S_{\text{max}}$ required to resolve the frequencies is given by
\begin{equation}
\label{iup}
  S_{\text{max}} \gtrsim 4\pi \bar\varrho(\omega)
\end{equation}
in terms of the local average density of frequencies $\bar\varrho(\omega)$.
Since this relation involves the average instead of the minimum spacing 
between frequencies, the signals can usually be significantly shorter 
than required by the Fourier transform.

Harmonic inversion of the signal (\ref{C_sc}) is a nontrivial task for the
following two reasons.
Firstly, the number of frequencies contained in the signal is usually large,
which implies that simple methods for solving the harmonic inversion problem
may be numerically unstable.
Secondly, the special functional form of a signal as a sum of $\delta$ 
functions does not fulfill the prerequisites of several established algorithms
that $C(s)$ should be known on an equidistant grid, $s=n\tau$, with 
$n=0,1,2,\dots$ \cite{Man97,Bel00}.
We briefly review the four numerical methods which so far have been 
successfully applied to the harmonic inversion of signals $C(s)$ given 
as a sum of $\delta$ functions.

\paragraph*{Method 1: Discrete FD ~---}
A powerful tool for the harmonic inversion of signals given on an equidistant
grid is the filter-diagonalization (FD) method of Wall and Neuhauser 
\cite{Wal95} in the version of Mandelshtam and Taylor \cite{Man97}.
The basic idea is to introduce a set of basis functions localized in 
frequency space and to recast the harmonic inversion problem as a 
generalized eigenvalue problem.
For a suitable choice of the frequency window the subset of frequencies 
located in that window can be obtained by solving a generalized eigenvalue 
equation with small matrices.

To evaluate the signal (\ref{C_sc}) on an equidistant grid, the $\delta$ 
functions must be given an artificial width $\sigma$ which spans several 
grid points, i.e., $\sigma > \tau$.
This regularization can be achieved by convoluting the signal with a 
narrow, e.g., Gaussian function.
In this form the FD method has been applied in Refs.\ \cite{Mai98,Mai99} as
a tool for semiclassical periodic orbit quantization.

The convolution of the signal $C(s)$ results in a damping of the amplitudes
$d_k \to d_k^{(\sigma)} = d_k \exp(-w_k^2\sigma^2/2)$.
The width $\sigma$ of the Gaussian function should be chosen sufficiently 
small to avoid an overly strong damping, e.g., by setting 
$\sigma \lesssim |w_{\text{max}}|^{-1}$ where $w_{\text{max}}$ is the 
largest frequency of interest.
To properly sample each Gaussian a dense grid with sufficiently small 
step size ($\tau\approx\sigma/3$) is required.
Therefore, the convoluted signal to be processed by FD usually consists of
a large number of data points, in particular when high frequency regions 
of the signal are to be analyzed.
The numerical treatment of such large data sets may suffer from rounding 
errors and loss of accuracy.

\paragraph*{Method 2: $\delta$-function FD ~---}
The artificial smoothing of the signal can be circumvented when following 
a different approach  suggested by Gr\'emaud and Delande \cite{Gre00}.
In contrast to Ref.\ \cite{Man97} where the signal is assumed to be known
on an equidistant grid, Gr\'emaud and Delande start from a continuous-time 
formulation of the FD algorithm close to the original ansatz in \cite{Wal95}.
Integrals involving the $\delta$ function signal (\ref{C_sc}) can then easily
be calculated and expressed in terms of periodic orbit sums.

\paragraph*{Method 3: Discrete DSD ~---}
An alternative method to FD for high-resolution signal processing is   
decimated signal diagonalization (DSD), which was introduced by Belki\'c 
{\em et al.}\ \cite{Bel00}.
DSD is a two-step process for harmonic inversion.
In the first step a low-resolution frequency filter is applied by 
subjecting the signal to a discrete Fourier transform, selecting the
Fourier components in the frequency interval of interest, and applying 
an inverse Fourier transform to them.
The resulting band-limited signal contains only a small number of 
frequencies in the chosen interval and can therefore be further processed,
in the second step, by conventional high-resolution methods such as, e.g.,
Linear Prediction or Pad\'e Approximants \cite{Mar87,NumRec}.
DSD effectively uses, in this processing stage, the operational part of
the discrete version of FD \cite{Man97}, which constructs small matrices 
of a generalized eigenvalue problem directly from digitized points of the
band-limited decimated signal.
The DSD technique is designed for signals given on an equidistant grid but 
can be applied to the $\delta$ function signal (\ref{C_sc}) after convolution
with a narrow, e.g., Gaussian function in the same way as explained above 
(see Method 1).

The DSD method of Ref.\ \cite{Bel00} is easy to implement as it 
basically resorts to standard algorithms for discrete Fourier transform 
and matrix diagonalization.
However, the application of the low-resolution Fourier filter in the first
step of the method implicitly assumes periodicity of the signal (with period
equal to the signal length), in which case the DSD filter is exact.
In general, of course, this condition is not met, so that only approximate
filtering can be achieved.
Therefore, DSD must be expected to be less accurate than FD (Method 1) for
frequencies close to the borders of the window, or when very short frequency
windows are chosen (see the discussion in Sec.\ \ref{results:sec}).

\paragraph*{Method 4: $\delta$-function DSD ~---}
The DSD technique (Method 3) can be modified for a more direct application
to the $\delta$ function signal (\ref{C_sc}) without the necessity of
convoluting the signal with a narrow, e.g., Gaussian function.
Because the signal $C(s)$ is given as a periodic orbit sum of $\delta$ 
functions the `filtering' step can be performed analytically by replacing
the discrete Fourier transform of Method 3 with the continuous form of
the Fourier transform and expressing the integrals in terms of periodic
orbit sums.
This technique was proposed in Ref.\ \cite{Mai00}.
The application of the analytical filter for a rectangular frequency window
$[w_0-\Delta w,w_0+\Delta w]$ results in a band-limited (bl) signal 
\cite{Mai00}
\begin{equation}
\label{C_bl}
  C_{\rm bl}(s) = \sum_{\rm po} {\cal A}_{\rm po}
     {\sin{[(s-s_{\rm po})\Delta w]}\over
     \pi(s-s_{\rm po})} \text{e}^{\text{i}s_{\rm po}w_0} \; ,
\end{equation}
where the $\delta$ functions are basically replaced with sinc functions
[${\rm sinc\ }x=(\sin x)/x$].
The band-limited signal (\ref{C_bl}) can be discretized with a small number
of points and further processed, in the second step, by conventional
harmonic inversion methods as described above (see Method 3).

In practice, the band-limited signal can only be evaluated approximately 
because the signal is only known up to a finite length.
Since the sinc functions decay slowly at infinity, peaks well beyond the end
of the known signal may have an influence on the band-limited signal points.
Omitting contributions from the (unknown) peaks beyond the limit of the
given signal amounts to the assumption that the signal be zero outside the
given range.
Note that this filter differs from the low-resolution filter of Method 3 
where the signal is implicitly assumed to be periodic.

In summary, the four methods  can be classified according to whether they are
discrete-time algorithms (Methods 1 and 3), which require a regularization
of $\delta$ function signals to be discretized, or continuous-time algorithms
adapted to $\delta$ function signals (Methods 2 and 4).
Alternatively, they can be classified into filter-diagonalization (FD)
methods (Methods 1 and 2) and decimated signal diagonalization (DSD) methods 
(Methods 3 and 4) where the low-resolution `filtering' and high-resolution 
signal processing is performed in two separate steps.

\section{Numerical examples and discussion}
\label{results:sec}
To quantitatively assess the relative performances of the different 
algorithms, we present a comparison of the numerical accuracy achieved 
by all of these methods for two simple but archetypal examples, viz.\ 
firstly, the high-resolution analysis of the spectrum of the harmonic 
oscillator and, secondly, the search for the zeros of Riemann's zeta function 
as a mathematical model for periodic orbit quantization in chaotic systems.
We choose these systems because they possess exact trace formulae, so that 
the decomposition of the semiclassical signal in a sum of exponentials is 
known to be exact.

\subsection{Harmonic oscillator}
The one-dimensional harmonic oscillator (with $\hbar\omega=1$) has energy 
eigenvalues $E_n=n+\frac 1 2$, $n=0,1,2,\dots$.
Its density of states can be written as an exact trace formula \cite{Bra97}:
\begin{equation}
\label{harm_osc:eq}
  g(E) = \sum_{n=0}^{\infty} \delta (E-E_n)
       = \sum_{k=-\infty}^{\infty} (-1)^k \text{e}^{2\pi \text{i} k E} \;.
\end{equation}
The right-hand side of Eq.\ (\ref{harm_osc:eq}) can be interpreted as a 
periodic orbit sum [in analogy to Eq.\ (\ref{spur})] where $S_k=2\pi kE$
is the action of the ($k$ times traversed) periodic orbit and $d_k=(-1)^k$
is the periodic orbit amplitude.
[The interpretation of the $k=0$ Thomas-Fermi term is special, see Ref.\
\cite{Bra97} for more details.]
The high-resolution analysis of the quantum spectrum
$g(E)=\sum_{n=0}^{\infty}\delta(E-E_n)$
should yield equally spaced real frequencies $\omega_k=2\pi k$ and amplitudes 
$d_k=(-1)^k$ of equal magnitude.
Thus, this simple application of harmonic inversion to the extraction of 
classical periodic orbit parameters from a quantum spectrum 
\cite{Mai99,Mai97,Gre00} is ideally suited to compare the efficiencies of the
different methods for the harmonic inversion of $\delta$ function signals.

Since the signal is periodic with period $\Delta E=1$, an integer signal 
length would render the low-resolution Fourier filter of Method 3 exact.
To avoid this atypical situation, we choose signal lengths as rational 
multiples of $\pi$.
According to the informational uncertainty principle (\ref{iup}) a signal 
length of $E_{\text{max}}\gtrsim 2$ should suffice to resolve the frequencies.
Typically, Eq.\ (\ref{iup}) slightly underestimates the required signal length.
We therefore present results calculated with a signal of length 
$E_{\text{max}}=\pi$, which means that only three energy levels contribute
to the signal.
To assess the accuracy of the results, we use the absolute values of the 
imaginary parts of the calculated frequencies and amplitudes as error 
indicators.
If the analysis of the signal were exact, all imaginary parts should vanish.
Therefore, an inspection of the sizes of the imaginary parts allows us to 
check the accuracy of the calculation.
We note that this sort of accuracy test can be applied to all bound systems.
If the exact frequencies are known, as is the case in our example systems,
the real parts can also be compared.
The errors of the real and imaginary parts typically are of the same order 
of magnitude and exhibit, at least qualitatively, the same behavior.

\begin{figure}
\vspace{9.0cm}
\includegraphics{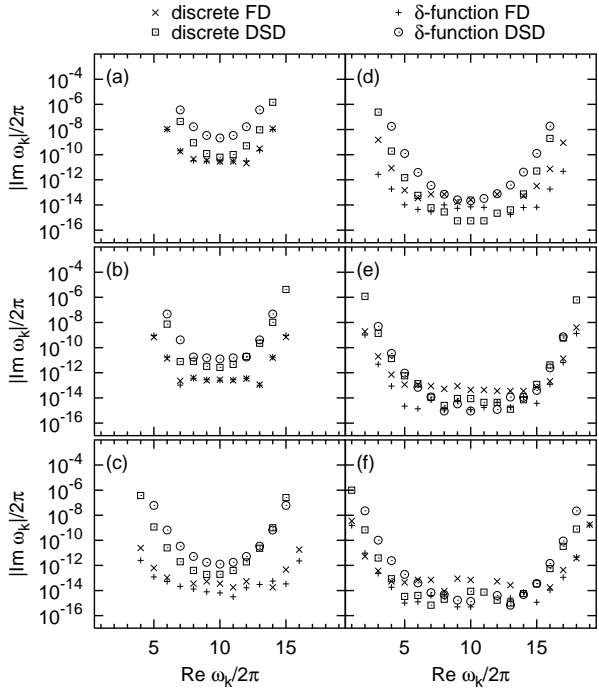}
\caption{Imaginary parts (absolute values) of the frequencies
         $\omega_k$ extracted from a harmonic oscillator signal of length
         $E_{\text{max}}=\pi$.
         Symbols $\times$, $+$, $\boxdot$, and $\odot$ denote to Methods
         1 to 4, respectively.
         Windows are $[10-\Delta \omega, 10+\Delta \omega]$ with 
         $\Delta \omega=$ (a) 4.5, (b) 5.5, (c) 6.5, (d) 7.5, (e) 8.5,
         (f) 9.5.}
\label{fig1}
\end{figure}

\begin{figure}
\vspace{9.0cm}
\includegraphics{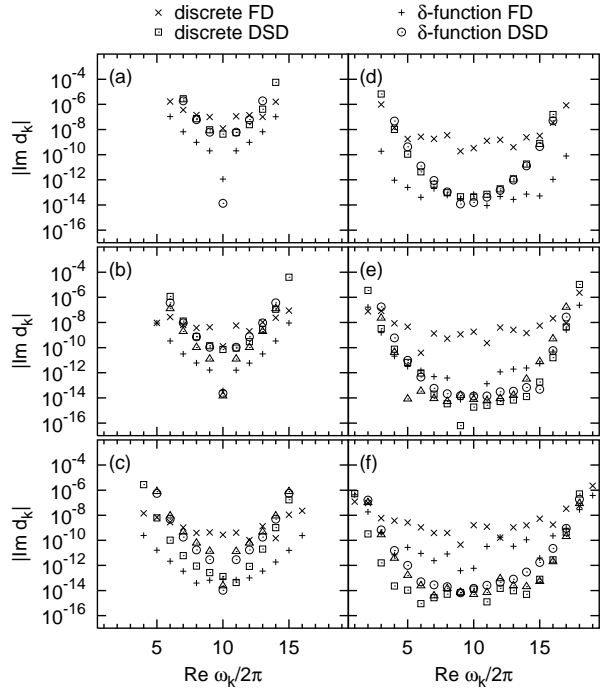}
\caption{Same as Fig.\ \ref{fig1} but for the imaginary parts of the 
         amplitudes $d_k$.}
\label{fig2}
\end{figure}

Results for the harmonic inversion of the quantum spectrum $g(E)$ obtained
with the four methods introduced in Sec.\ \ref{methods:sec} are presented in
Figs.\ \ref{fig1} and \ref{fig2} for the imaginary parts of the frequencies 
$\omega_k$ and amplitudes $d_k$, respectively.
For frequencies which appear to be missing, imaginary parts of zero have been
obtained by the pertinent method.
From figure parts (a) to (f) the width $\Delta\omega$ of the frequency filter 
is increased.
For the application of Methods 1 and 3 the signal has been discretized with
step width $\tau=0.002$ after convolution of the signal with a Gaussian 
function with width $\sigma=0.006$.
In all cases it can be seen that the precision achieved decreases to the 
boundaries of the frequency window.
The reason is that none of the filtering methods is exact and can neither 
completely remove the influence of frequencies outside the window of interest
nor exactly preserve the contributions of frequencies inside the window.
For narrow windows, the FD methods 1 and 2 outperform the DSD algorithms 3 
and 4, for wide windows the situation is reversed.
The frequencies obtained by Methods 1 and 2 are equally precise for small 
windows, whereas for wide windows Method 2 gains superiority and even 
competes with the DSD methods.
In general, the distance from the 
\newpage
\phantom{}
\noindent
window boundaries where a method acquires its full 
precision is smaller for the FD than for the DSD methods.
Calculations were carried out with double precision.
For the widest window shown, frequencies have practically been obtained 
to machine precision.

For all methods, the precision of the amplitudes in Fig.\ \ref{fig2} is 
somewhat lower than that of the frequencies in Fig.\ \ref{fig1} because 
the amplitudes are calculated from the eigenvectors of a generalized
eigenvalue problem, which in general are less accurate than the eigenvalues.
In particular, the difference in precision between the frequencies and 
amplitudes is considerably larger for Method 1 than for any other method,
so that even for small windows the amplitudes obtained by this technique 
are the least accurate (see the $\times$ symbols in Fig.\ \ref{fig2}).

\subsection{Zeros of Riemann's zeta function}
It is a peculiar feature of the harmonic oscillator signal that the
density of frequencies is constant, i.e., the precision of frequencies 
obtained from a signal of a given length is the same throughout the 
whole frequency domain.
However, in typical systems the density of states grows rapidly with
the frequency, which means that a longer signal is required to extract 
higher frequencies.
As an example of a system exhibiting this typical behavior, we discuss 
the Riemann zeta function which has served 
\newpage
\phantom{}
\noindent
as a mathematical model for periodic orbit quantization \cite{Mai98,Ber86}. 
It is well known that,
if the zeros of $\zeta(z)$ on the critical line $\text{Re\ } z=\frac 1 2$ 
are written as $z =\frac 1 2 - \text{i} w$, the density of zeros on the 
critical line can be expressed as \cite{Ber86}
\begin{equation}
\label{g_zeta}
  g(w) = - \frac 1 \pi \sum_p \sum_{m=1}^\infty \frac{\ln p}{p^{m/2}}
         \cos(wm\ln p) \; ,
\end{equation}
where $p$ runs over all prime numbers.
Eq.\ (\ref{g_zeta}) is formally identical to a semiclassical trace formula 
with $S_{pm}=wm\ln p$ corresponding to classical actions and 
${\cal A}_{pm}=(\ln p)/p^{m/2}$ corresponding to classical amplitudes.
With this interpretation, the Riemann zeta function can be used as a 
mathematical model for chaotic dynamical systems, and the Riemann zeros are 
obtained by harmonic inversion of the $\delta$ function signal \cite{Mai98}
\begin{equation}
\label{C_zeta}
  C(s) = \sum_p \sum_{m=1}^\infty \frac{\ln p}{p^{m/2}} \delta(s-m\ln p) \; .
\end{equation}
Unlike typical semiclassical trace formulae, Eq.\ (\ref{g_zeta}) is exact.
As the zeta function has only simple zeros, the amplitudes $d_k$ obtained 
from the harmonic inversion of the signal (\ref{C_zeta}) must all be equal 
to 1.

\begin{figure}
\vspace{9.0cm}
\includegraphics{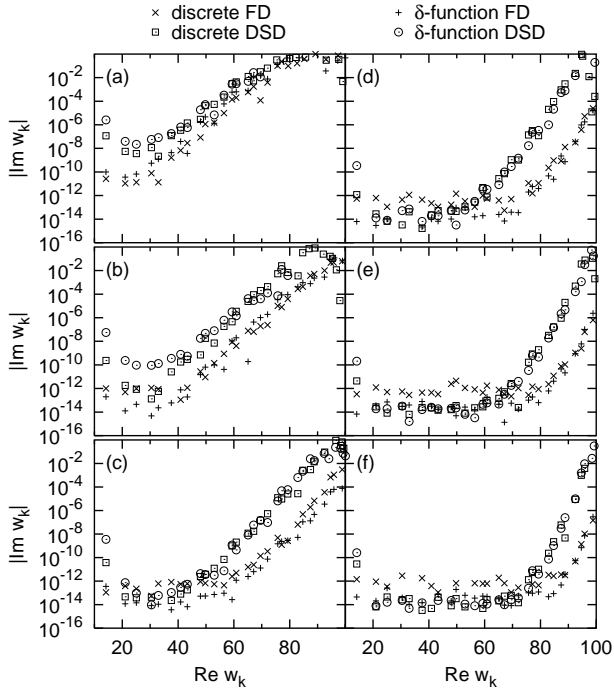}
\caption{Imaginary parts (absolute values) of locations $w_k$ of zeros
         of the Riemann zeta function in the frequency window $[10,100]$. 
         Symbols $\times$, $+$, $\boxdot$, and $\odot$ denote to Methods
         1 to 4, respectively.
	 Signal lengths are $S_{\text{max}}=$
         (a) 4.5, (b) 5.0, (c) 5.5, (d) 6.0, (e) 6.5, (f) 7.0.}
\label{fig3}
\end{figure}

In Fig.\ \ref{fig3} we present our numerical results obtained by application
of Methods 1 to 4  to extract the 
(numerically complex valued) Riemann zeros with $\text{Re}\ w_k<100$.
Ideally, all values $w_k$ should be real.
Therefore, the absolute values of the imaginary parts of the $w_k$ can again
serve 
\newpage
\phantom{}
\noindent
as a measure for the accuracy of the harmonic inversion process.
For the application of Methods 1 and 3 the signal has been discretized with
step width $\tau=0.002$ after convolution of the signal with a
Gaussian function with width $\sigma=0.006$.

It is no problem to construct the signal (\ref{C_zeta}) for the Riemann zeros
up to quite large maximum values $S_{\text{max}}$ because only prime numbers 
are used as input.
However, the periodic orbit quantization of physical systems usually requires
a numerical periodic orbit search which becomes more and more expensive for
longer orbits, especially in chaotic systems, where the number of orbits
proliferates exponentially with increasing signal length.
Therefore, in practical periodic orbit quantizations the given signal length
is often rather short.
In Fig.\ \ref{fig3} we present the results for the accuracy of the 
methods for harmonic inversion for various signal lengths, increasing from
$S_{\text{max}}=4.5$ in Fig.\ \ref{fig3}a to $S_{\text{max}}=7.0$ in Fig.\ 
\ref{fig3}f.
The frequency window $w\in [10,100]$ is kept fixed.

For a fixed signal length, the zeros up to a certain critical value can 
be obtained to a constant precision.
Above the critical frequency, the precision decreases rapidly due to the 
higher density of states.
As was to be expected, for all methods the critical frequency increases 
with growing signal length, which means that frequencies in regions of 
higher spectral density can be resolved.
Roughly, the critical frequency is determined by the informational 
uncertainty principle (\ref{iup}).
In fact, it is slightly higher for
\newpage
\phantom{}
\begin{figure}
\vspace{9.0cm}
\includegraphics{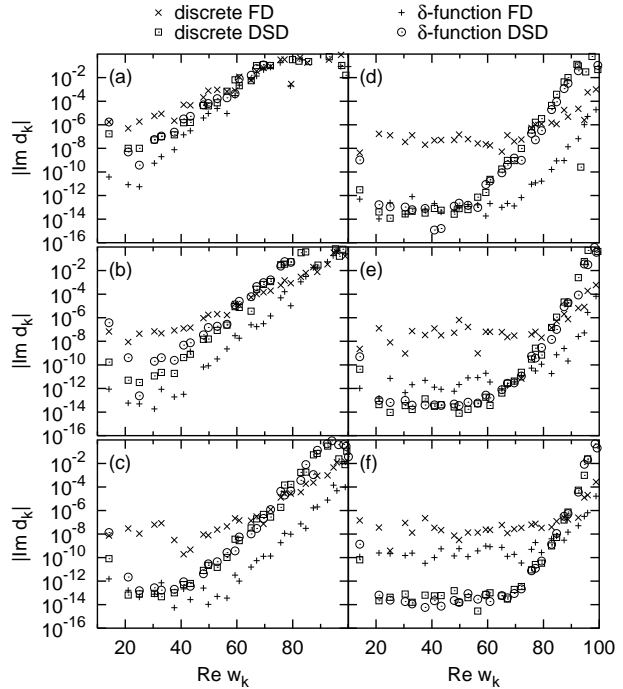}
\caption{Same as Fig.\ \ref{fig3} but for the imaginary parts of the
         multiplicities $d_k$.}
\label{fig4}
\end{figure}
\noindent
the FD methods 1 and 2 than for the DSD methods 3 and 4.
As before, the maximum accuracy below the critical frequency is obtained 
by the DSD methods.
However, above the critical frequency the precision yielded by the FD
methods is higher.

The lowest zero of the zeta function is located at $w=14.1347$, not far 
above the lower boundary of the frequency window at $w=10$. 
For the first zeros a decrease in accuracy due to the proximity of the 
boundary can be seen.
Evidently, the influence of the boundary diminishes with increasing signal 
length.
Again, it is considerably more pronounced for the DSD than for the FD methods.
For the latter, it can only be seen in the shortest signals.
With any of the four methods, the boundary effects on the lowest zeros can be
removed if the lower boundary of the frequency window is decreased to $w=0$.

Fig.\ \ref{fig4} presents results similar to those shown in Fig.\ \ref{fig3}, 
but for the imaginary parts of the multiplicities $d_k$.
The accuracy of the results achieved with the different methods resemble 
those obtained for the imaginary parts of the frequencies $w_k$, with the 
exception of Method 1 (see the $\times$ symbols) which provides amplitudes 
with significantly lower precision.

\section{Conclusion}
In this paper we have quantitatively determined the accuracies of four 
different algorithms for the high-resolution harmonic inversion of $\delta$ 
function signals, by applying all algorithms to two, physically motivated, 
example signals. 
For sufficiently long signals and broad frequency windows the four methods
provide excellent results of very high accuracy, in the case of the 
examples selected even close to machine precision.
However, when either the width of the frequency filter or the signal length
is considerably reduced, the accuracy of the results obtained by the four 
methods can vary by several orders of magnitude.

Our calculations show that no general clear-cut answer to the question
``Which method is best in all physical situations?'' is possible.
In practice, the window width can be regarded as a free parameter, i.e., 
it can usually be chosen sufficiently large to achieve good results before 
increasing computational effort or numerical instabilities become a
restriction.
The signal length, on the contrary, is often fixed or at least hard to 
increase.
In such a case the choice of the algorithm for harmonic inversion of the
signal will be essential to achieve the optimum results.
When the signal length is quite at the limit for convergence of the 
frequencies and amplitudes, the filter-diagonalization (FD) methods 1 and 2 
provide superior accuracy compared to the decimated signal diagonalization 
(DSD) methods 3 and 4.
For signals given as the sum of $\delta$ functions Method 2 will often 
prove to be the method of choice. 

We conclude by noting that harmonic inversion techniques can be generalized
so as to cope with the analysis also of multidimensional signals, with 
important applications in other areas of physics \cite{Man00}.
The knowledge gained from the comparison of methods for one-dimensional 
harmonic inversion in this paper should also serve as a useful guide in 
future developments and applications of accurate and efficient algorithms 
for multidimensional high-resolution signal processing.

\acknowledgements
We thank P. A. Dando and H. S. Taylor for fruitful discussions.
This work was supported by the Deutsche For\-schungs\-ge\-mein\-schaft 
and Deutscher Akademischer Aus\-tausch\-dienst.

\end{document}